\documentclass[a4paper,11pt]{article}
\usepackage{pos}
\usepackage{comment}
\usepackage{siunitx}
\usepackage{subcaption}
\usepackage{lineno}
\title{Concept Study of a Radio Array Embedded in a Deep Gen2-like Optical Array}

\author{The IceCube Gen2 Collaboration \\{\normalsize \normalfont(a complete list of authors can be found at the end of the proceedings)}}

\emailAdd{abigail.bishop@wisc.edu}

\abstract{

The IceCube Neutrino Observatory has discovered a diffuse astrophysical flux up to $\SI{10}{PeV}$ and is now planning a large extension with IceCube-Gen2, including an optical array and a large radio array at shallow depth \cite{Gen2}. Neutrino searches for energies >$\SI{100}{PeV}$ are best done with such shallow radio detectors like the Askaryan Radio Array (ARA) or similar (buried as deep as 200 meters below the surface) as they are cheaper to deploy. This poster explores the potential of opportunistically burying radio antennas within the planned IceCube-Gen2 detector volume (between 1350 meters and 2600 meters below the surface). A hybrid detection of events in optical and radio could substantially improve the uncertainty of neutrino cascade direction as radio signals do not scatter in ice. We show the first results of simulating neutrinos from an astrophysical and a cosmogenic flux interacting with 9760
ARA-style vertically polarized radio antennas distributed evenly across 122
strings.

\vspace{4mm}
{\bfseries Corresponding authors:}
Abby Bishop$^{1*}$, Lu Lu$^{1}$, Albrecht Karle$^{1}$, Ben Hokanson-Fasig$^{1}$\\
{$^{1}$ \itshape University of Wisconsin - Madison}\\[4mm]
$^*$ Presenter

\FullConference{37$^{\rm{th}}$ International Cosmic Ray Conference (ICRC 2021)\\
		July 12th -- 23rd, 2021\\
		Online -- Berlin, Germany}

}

\begin{document}
\maketitle

\section{Introduction}\label{sec:intro}

    
    IceCube has detected a diffuse cosmic neutrino flux in the energy range from $\SI{60}{TeV}$ to more than $\SI{10}{PeV}$ \cite{ICFlux1,ICFlux2,ICFlux3,ICFlux4}.  
    IceCube-Gen2 will substantially improve the event rates especially for cascade events. 
    IceCube-Gen2 will include a deep optical array of about 8 km$^3$ instrumented volume and the Gen2-Radio array for high energy (>$\SI{30}{PeV}$) neutrino observations \cite{Gen2}. 
    The Gen2-Radio array is designed using insights from previous radio neutrino experiments such as RICE \cite{RICE}, ANITA \cite{ANITA}, ARA \cite{ARA, ARA2}, ARIANNA \cite{ARIANNA1, ARIANNA2}, and RNOG (currently in construction in Greenland) \cite{RNOG}.
    The Gen2-radio array will achieve a sensitivity at high energies (above 30 PeV)
    cosmic neutrino flux by more than an order of magnitude of what would be imaginable with an optical array.  
Here, we are investigating the possibility to enhance the performance of 
the Gen2 optical array in the range from a few PeV to above 100 PeV
by co-deploying radio antennas together with the optical sensors, one antenna 
for each optical module.  
In this energy range, Gen2 will over the time space of a decade, observe more than 50 neutrino induced cascade events  above 10 PeV. 
Useful experiences exist from the deployment of AURA
(Askaryan Under Ice Radio Array, \cite{AURA} a prototype radio detector 
that was co-deployed with IceCube.  

    Could the additional radio data improve Gen2-Optical measurements or help connect flux measurements between $\SI{3}{PeV}$ and $\SI{100}{PeV}$?  
    The considered radio component is not a stand alone detector for detecting ultrahigh energy neutrinos.  It primarily would be an augmentation of optical array
    which would provide all triggering.  
    
    The results indicate that a substantial amount of instrumentation would be required. Initial considerations suggest that the implementation in the Gen2 architecture would be quite straightforward.  The basic configuration is that
    one antenna would be hooked up with every optical module.  
    The instrumentation showcased here is not planned for Gen2, however the reader may find characteristics of such a configuration, 
    studied here for the first time, to be intriguing.  

\section{Simulation and detector configuration} 
  The simulation package "Python for Radio Experiments" (PyREx)\footnote{Github: https://github.com/bhokansonfasig/pyrex} was used to simulate 40,000 neutrino events at select  energies ranging from $\SI{1}{PeV}$ to $\SI{10}{EeV}$. 
  Neutrino interactions are simulated using the CTW model described in \cite{CTW-interaction}.
  The following neutrino properties are determined randomly according to the cosmogenic model: every flavor is equally possible and the probability that a particle is a neutrino (as opposed to an anti-neutrino) is 0.78, 0.61, and 0.61 for electron-, muon- and tau neutrino respectively \cite{Gandhi}. 
  After interacting, the simulation propagates Askaryan signals to all antennas following the Alvarez-Muniz, Romero-Wolf, Zas parameterization then simulates their responses \cite{ARZ-signal}.
  
  
  
  The radio detector component is conceived such that 
that every optical sensor on the string, there is one radio antenna associated 
with that sensor and located 8\,m above it.
  Thus, a simulation was performed with 9760 ARA-style vertically polarized (Vpol) radio antennas located on the 120 Gen2 Strings, 80 per string, between 1350 and 2600 meters below the Antarctic ice's surface.  The simulation assumed a vertical spacing of antennas is 17\,m and the spacing between strings is $\approx$~240\,m.  
  The technical implementation would be such that every multi-PMT optical sensor would feature one additional readout channel that would connect an envelope detector to the antenna, which is located $\SI{7}{m}$ above the DOM in the null of the antenna. 
  
  In the simulation 40,000 events are generated and simulated at select energies 
  using the radio-signal simulation software PyREx and the aforementioned parameters. Next, all events with $\SI{6}{PeV}$, $\SI{10}{PeV}$, $\SI{30}{PeV}$, or $\SI{100}{PeV}$ of energy and located within 300 meters of IceCube-Gen2 boundary are isolated and processed using the optical-signal simulation software IceTray\footnote{Github: https://github.com/icecube/icetray/tree/gen2-optical-sim} ($\approx$ 6,000 events for each energy).
  We primarily investigate a mode of operation in which
  the triggering would be provided entirely by the optical sensors. Thus, the Gen2 optical sensors would provide both the trigger for the readout and the readout channel for the RF signal. It would be realized as an additional channel to the 16 or 18 PMT channels in consideration for Gen2 \cite{ICRC_LOM}.  This trigger and readout would be quite straightforward to implement as the formation of the optical trigger is managed with the Gen2 optical trigger system.
  The readout decision is made at the DOM level and does not require a global event trigger   formation. 
  

Considerations of RF noise from nearby optical sensors will be ignored for this conceptual study, but would need to be considered if this topic is studied further. Based on previous measurements with the AURA detectors \cite{AURA}, it is assumed this would not be a major issue.

  \subsection{Optical-Radio Hybrid Events}
  
    Radio waves travel essentially unscattered in ice, unlike the optical signal.
    Therefore the radio signals could substantially improve angular reconstruction of events, especially for cascade-like events where the angular resolution of the optical detector is more limited. In some cases, it could also add substantial information to events with the vertex outside of the array. 
    The primary energy range of events in consideration is from a few PeV to $\SI{100}{PeV}$ - events with bright optical signals in the optical detector.
    
\begin{figure}[ht]
\begin{subfigure}{0.5\textwidth}
  \centering
  \includegraphics[width=1\linewidth]{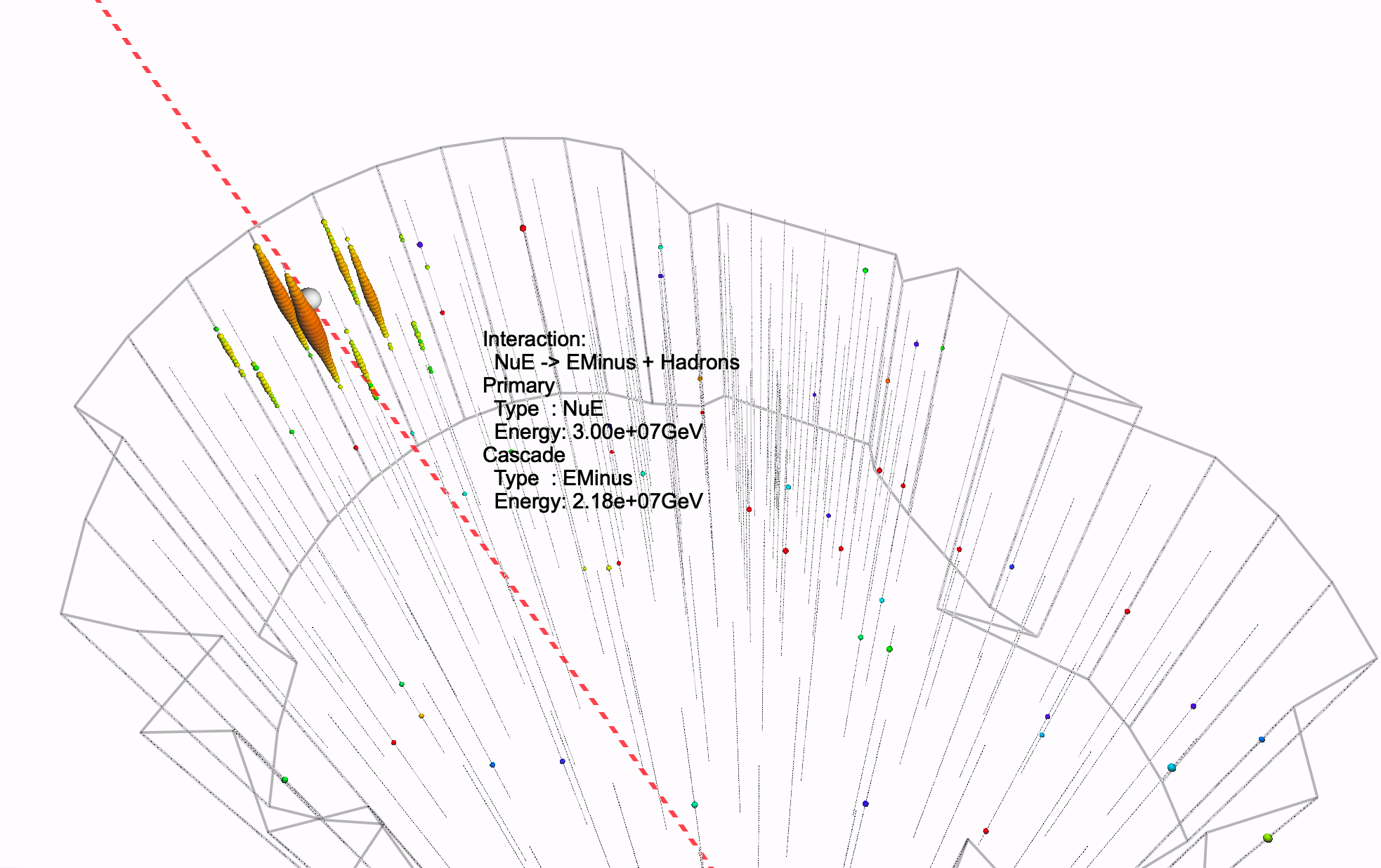}  
  \label{fig:eventview_optical}
\end{subfigure}
\begin{subfigure}{0.5\textwidth}
  \centering
  \includegraphics[width=1\linewidth]{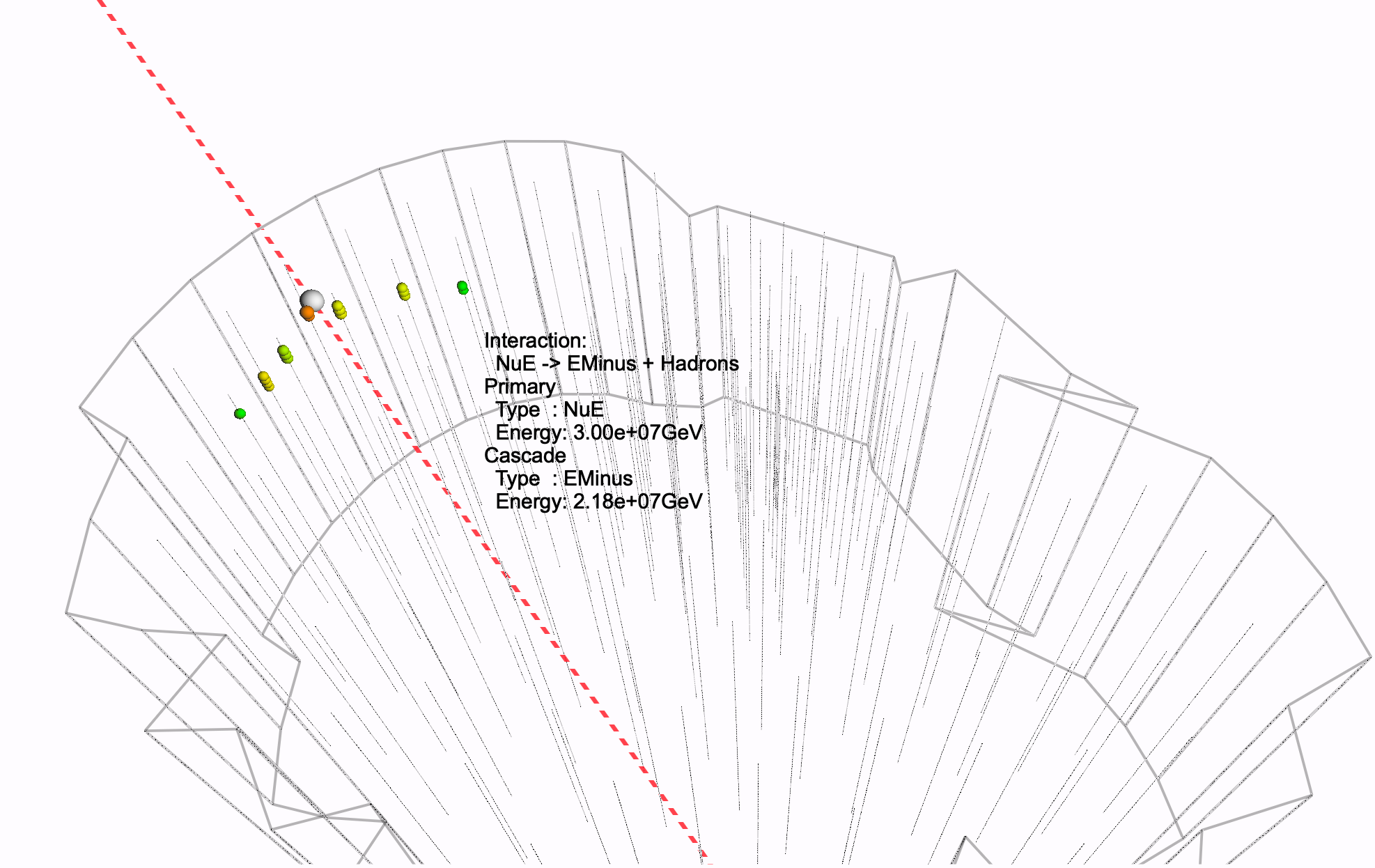}  
  \label{fig:eventview_radio}
\end{subfigure}
\caption{Image of the detector response by the Gen2-Optical detector and the hypothetical radio detector for a neutrino event that can be observed perfectly by both detectors. The size of each bubble corresponds to the strength of the received signal at the corresponding radio antenna or DOM, while the color corresponds to the time the signal peaked.}
\label{fig:goldenEvent}
\end{figure}

    To investigate a possible coincident detection of radio and optical signals, we start with the detailed examination of an event of $\SI{30}{PeV}$ energy and travelling downwards at a zenith angle of 65 degrees. 
    The event is simulated in both PyREx and IceCube's Gen2-Optical IceTray simulation to simulate both the radio and the optical response. 
    The detector response can be seen in Figure~\ref{fig:goldenEvent} where the 
    the detected event topology is shown for optical in the left panel and for RF in the right panel
    where the cone of coherent emission is clearly visible.  
    
\begin{figure}[ht]
\begin{subfigure}{0.42\textwidth}
  \centering
  \includegraphics[width=0.8\linewidth]{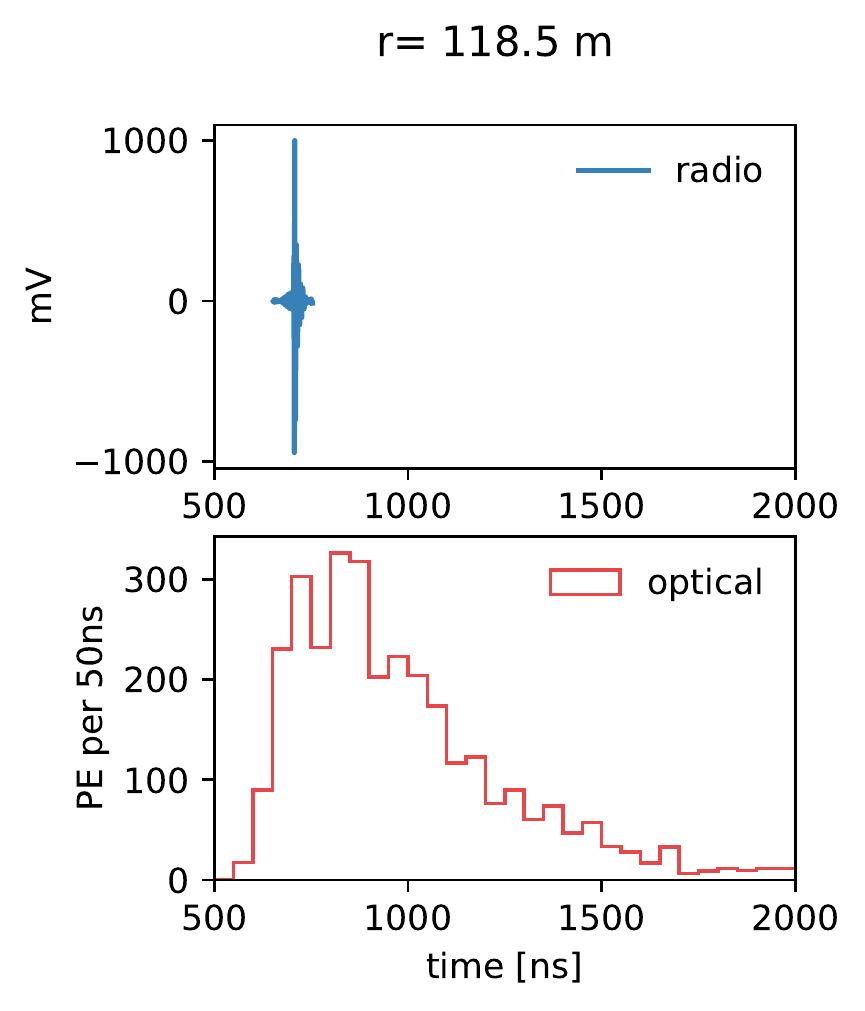}  
\end{subfigure}
\begin{subfigure}{0.58\textwidth}
  \centering
  \includegraphics[width=1\linewidth]{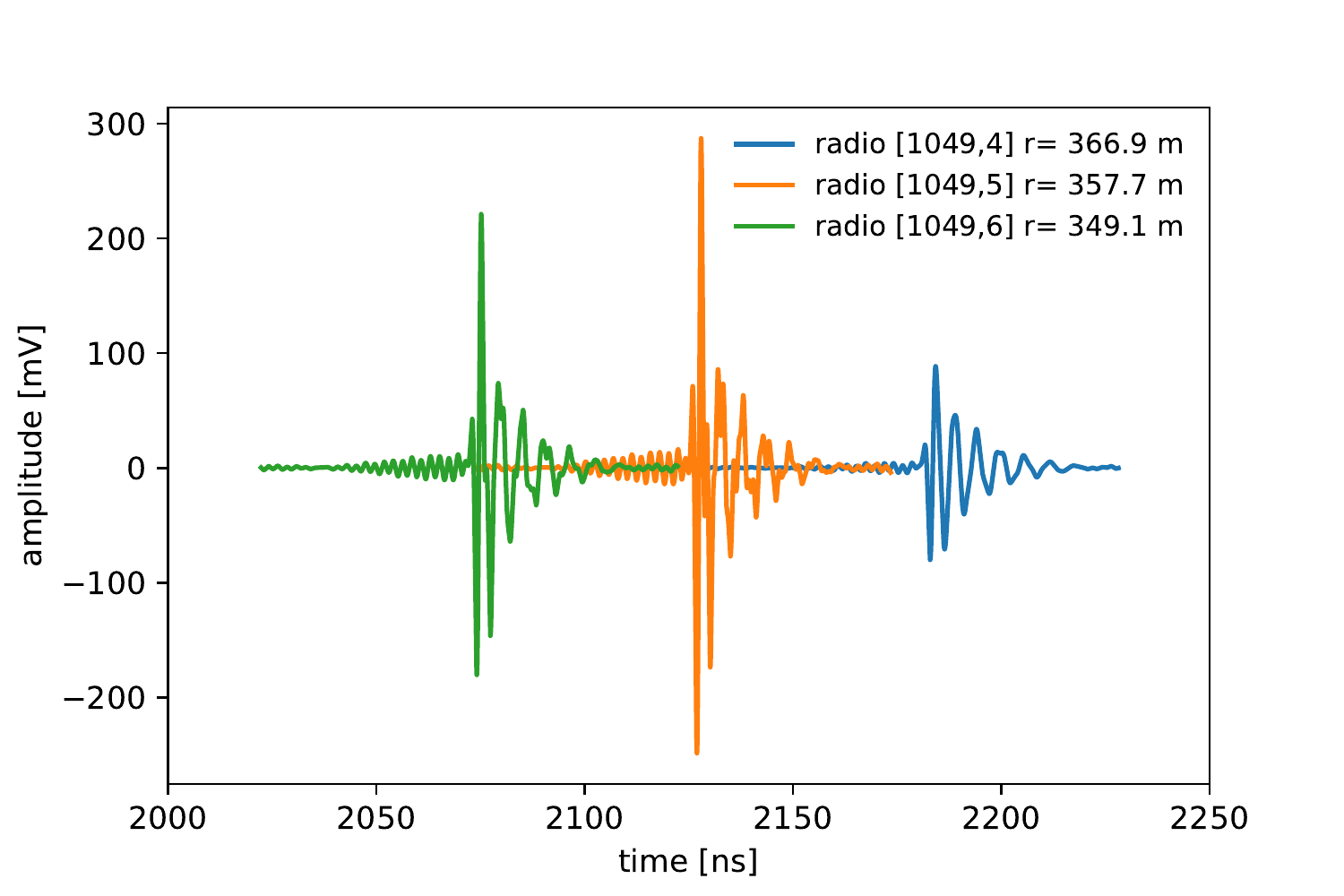}  
\end{subfigure}
\caption{The simulated optical and RF 
      waveforms are shown for the same neutrino event on the same time scale.  
      The optical sensor with the RF extension is located 
      at a distance of 118.5\,m from the vertex.  In the right panel three waveforms are shown for a string at a distance of 350\,m.}
\label{fig:goldenEvent-waveforms}
\end{figure}

    The optical and RF signal waveforms are shown in Figure~\ref{fig:goldenEvent-waveforms} for one location. 
    The  short impulsive radio signal is shown on the same time scale as the optical signal which extends over several 100\,ns. The time resolution of radio signal detection would be at the level of a few ns, limited only the system clock synchronization. The optical trigger for every module would be at the level of a few photo-electrons.
    As mentioned, the readout of every antenna would be initiated and performed by the local DOM.  
    In the right panel we show three waveforms recorded on consecutive antennas 
    on a string at a substantial distance of 350 to 367\,m. The vertical spacing of the antennas is 16\,m. The amplitude profile of the three signals will provide very good tool to 
    locate the intersection of the radio cone with the Gen2-string
    to a precision small compared to the antenna spacing.  

 \subsection{Event reconstruction}
        Although angular reconstruction techniques are still in-progress, the benefits of radio detectors locating these neutrinos are easy to picture from the separation of the radio and optical signals and the informative topology of the radio event, especially for cascades.
    The waveforms of photons from a PeV event in the optical extend typically for hundreds of nanoseconds, while the RF time resolution would be a few  nanoseconds.  
    
    It is important to recall that every event that will be considered will have a very bright optical signal of 
    typically more than 100,000 photoelectrons.  Thus the optical reconstruction will provide a good constraint on the vertex and a reasonable prior on the direction. Thus even a few relatively faint radio signals can  contribute significantly to the reconstruction.  
    For this reason, we consider a signal above 2$\sigma$ in at least 2 antennas 
    to be a threshold for hybrid events.  
    
\begin{figure}[ht]
    \begin{subfigure}{0.5\textwidth}
      \centering
      \includegraphics[width=1\linewidth]{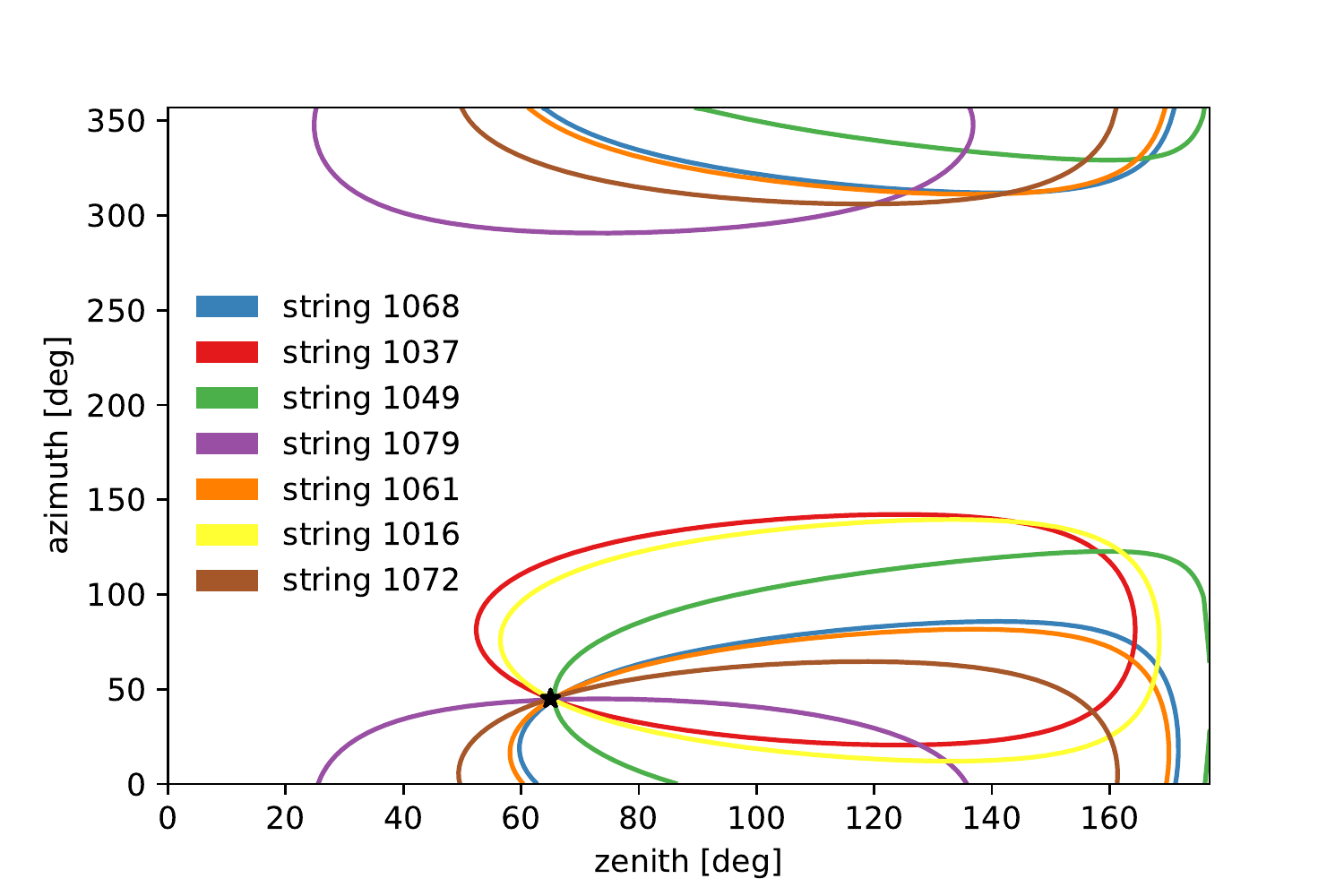}  
      \label{fig:goldenEventReco-zoomedout}
    \end{subfigure}
    \begin{subfigure}{0.5\textwidth}
      \centering
      \includegraphics[width=0.96\linewidth]{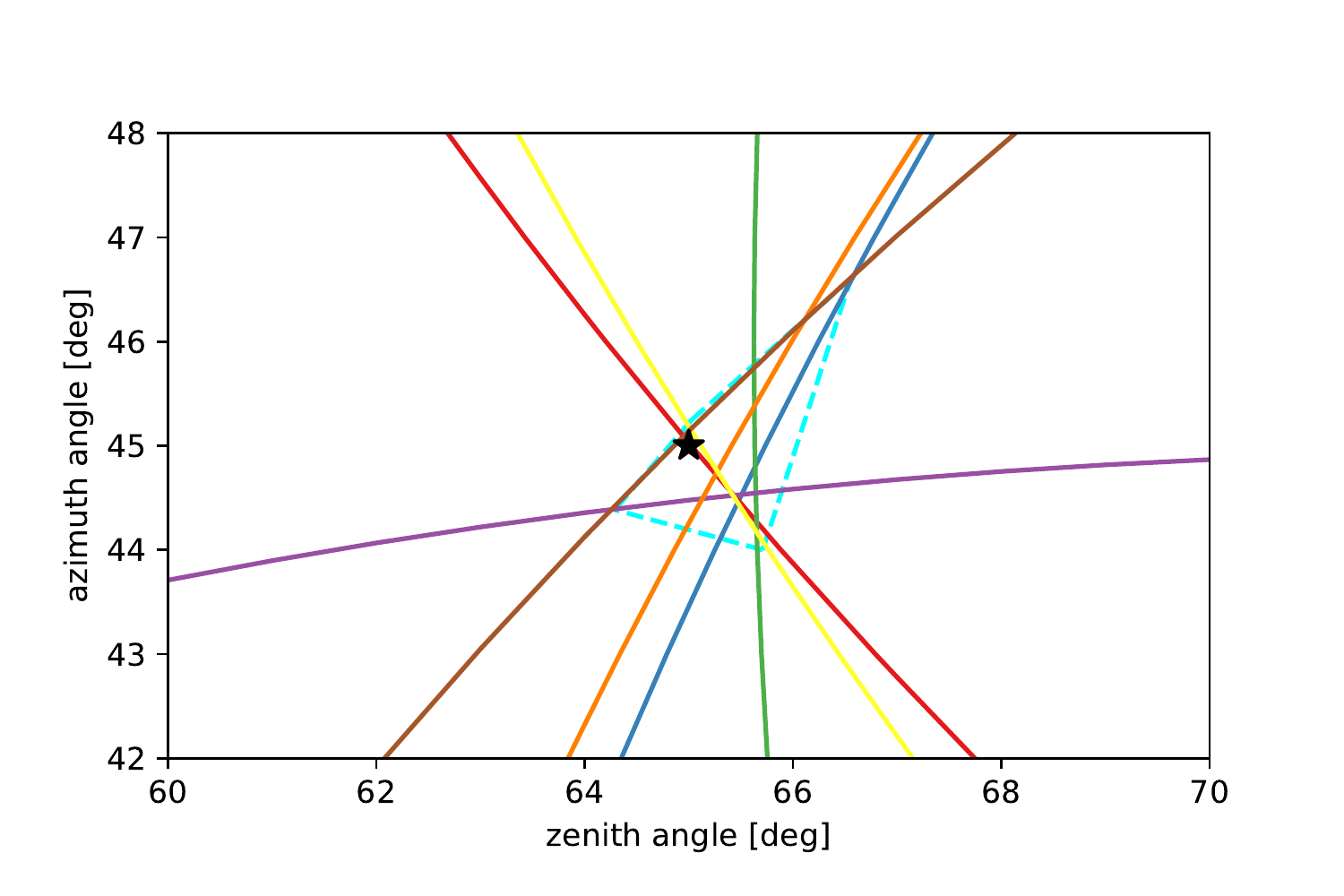}  
      \label{fig:goldenEventReco-zoomedin}
    \end{subfigure}
    \caption{A simple event reconstruction leads to one contour for every antenna signal.  Here the solutions for the brightest antenna signal on each string is shown. The intersection indicates the solution.
    The close-up in the right panel indicates a convex hull of the intersections, which may be considered as an estimate of the uncertainty of 
    less than 1 degree. 
    The true position of the event is: zenith=$65^\circ$  and azimuth=$45^\circ$. 
    The resolution would significantly improve the Gen2 optical cascade reconstruction 
    with a requirement of 10 degrees. }
    \label{fig:goldenEventReco}
    \end{figure}

   One strategy of event reconstruction uses the recorded RF signals to locate the  Radio Cherenkov cone.
   In this reconstruction we use the vertex as a given.
   The vertex can be determined in the optical channel to a precision 
   of about 20m.  The RF channel will allow a better vertex position of easily 10m. 
   However the timing signals do not provide the direction. 
   Instead we use the presence of RF signals as an indicator for the 
   proximity of the Cherenkov cone.  In a most basic version, we consider 
  the brightest antenna signal on a string to be exactly on cone.  
 Every antenna with a hit provides a constraint to the neutrino direction
 as illustrated in Figure \ref{fig:goldenEventReco}.
In the case of the simulated 30 PeV cascade there is a signal on 7 strings,
each of which provides one contour.  
The intersection of the contours indicates the direction of the event,
and the differences of the intersections indicate a measure of the uncertainty. 
The convex hull points to a resolution of better than 1 degree.
Simple optimizations of this reconstruction would include to 
use the amplitude of the antenna signals and their 
distance from the vertex as a weight in this procedure.
For example, here we used only the waveform with the largest amplitude
in Figure \ref{fig:goldenEvent-waveforms}. 
An obvious improvement would be fit the amplitude profile of the Askaryan cone, 
or simply take some weighted average as a measure of the RF Cherenkov cone.  
For an event like the one investigated a resolution of a fraction of a degree
is well conceivable.

  \subsection{Event rates}
    
   Since the focus of this investigation is on cascade events with possible RF content this  analysis  is based on simulations of electron and electron-anti neutrinos. The threshold for 
    the optical channel is set sufficiently high to include only bright events at PeV energy scale. 
    Every event at the energies considered (6, 10, 30 and $\SI{100}{PeV}$) will 
    produce a very bright signal in the optical detector.

    \begin{figure}
      \centering
    \includegraphics[width=4.5in]{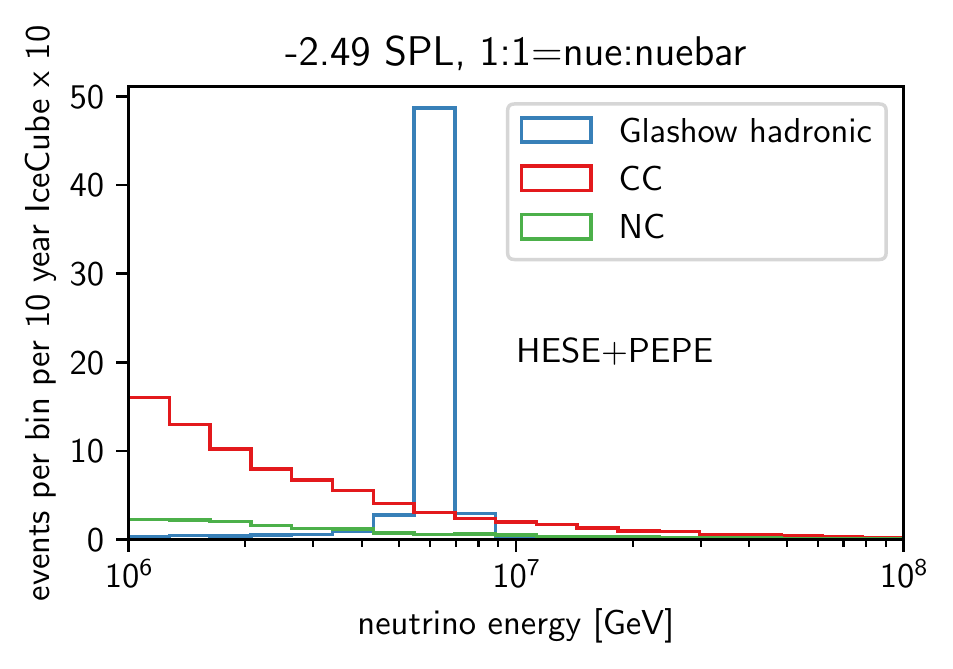}     
      \caption{Number of Glashow hadronic, charged current and neutral current events from electron (anti)neutrinos selected by HESE \cite{ICFlux1} and PEPE \cite{ICFlux4} analyses assuming a single power law flux  of E$^{-2.49}$ for 10 years of IceCube livetime. Numbers are multiplied by ten to scale up to the Gen2 event rate (which is approximately 10 times larger in instrumented volume than IceCube). }
      \label{fig:nuRate-Gen2}
    \end{figure}

    
     
   Figure \ref{fig:nuRate-Gen2} shows event rates for an injected
   IceCube neutrino flux of E$^{-2.49}$ using high energy starting event (HESE) \cite{ICFlux1} and PeV energy partially contained events (PEPE) data \cite{ICFlux4}.  
   In this figure, we assume and effective volume for detection cascade events to be 10
   times that of IceCube, in line with the ratio of the instrumented volumes of Gen2 and IceCube.
   For a livetime of 10 years 
   we obtain a total of 70 cascade events in the neutrino energy range from $\SI{5}{PeV}$ to $\SI{100}{PeV}$, which includes an important contribution of 54 events from the Glashow resonance.
 This channel is especially intriguing, as it will 
 provide an ensemble of electron antineutrinos at sub-degree resolution. 
    We then ask the question, what is the fraction of events that is seen in the optical that will also be seen in RF.

\begin{figure}
      \centering
      \includegraphics[width=4.5in]{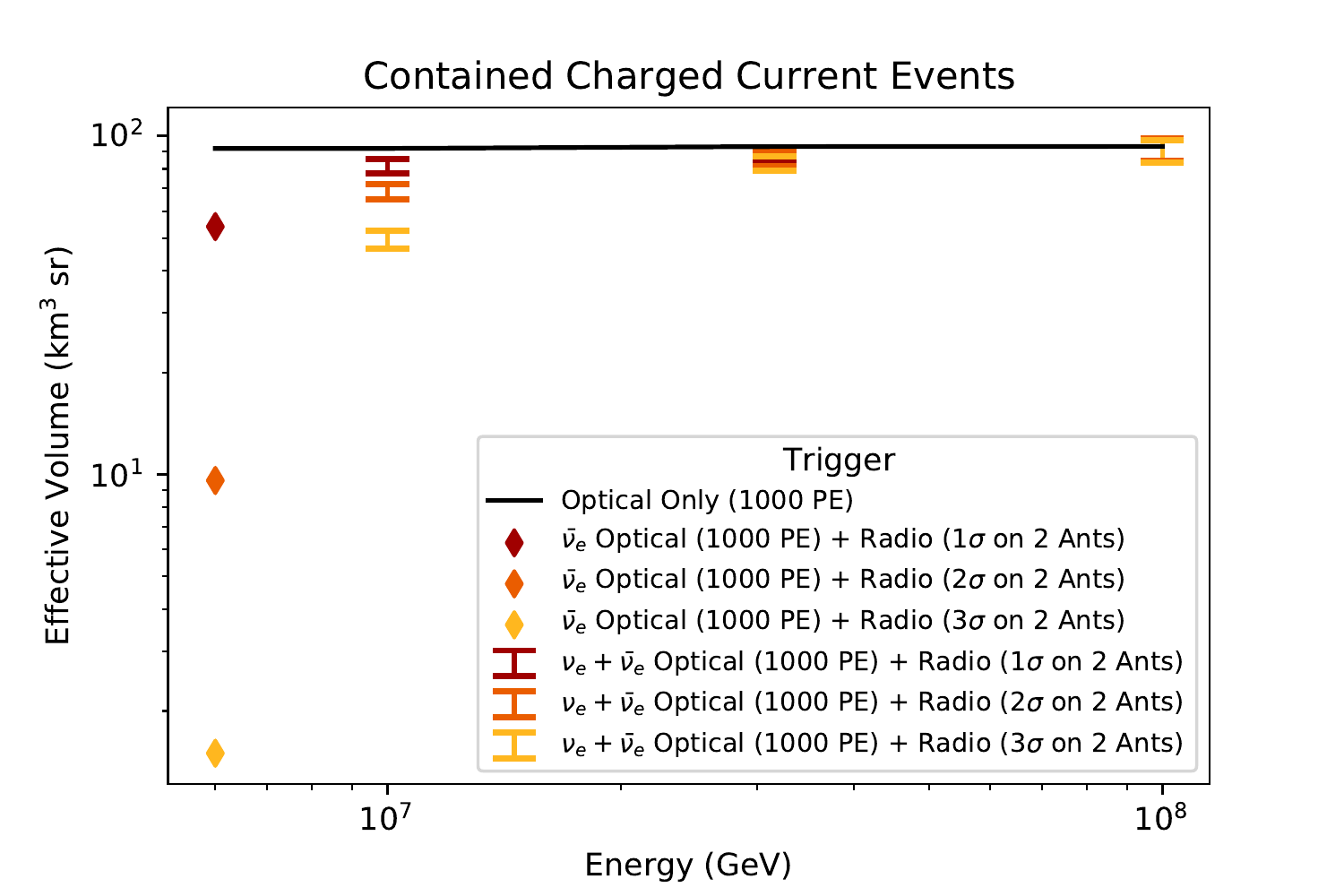}
      \caption{Effective volumes are shown for electron neutrino induced cascade events that trigger both Gen2-Optical and have a signal in the hypothetical radio detector for varying radio trigger requirements.
      Also indicated is a the geometric volume of the Gen2 optical 
      array that is fully effective for the set condition at all energies. }
      \label{fig:veffs-hybrid}
    \end{figure}

      For that we determine the effective volume of such hybrid radio and optical configuration with signals in the RF channel.  The effective volume may be seen in Figure~\ref{fig:veffs-hybrid}
      for three threshold conditions  in the RF channels.
   One could imagine to use antennas more optimized for the purpose and taking 
     advantage of the larger hole size.  A gain in effective height 
     of the antenna of a factor of 2 or 3 seems plausible.  This is in part
     the reason why we show results with several thresholds, including 2 and 1 sigma on 2 antennas.  
     For illustration we show the effective volume for three threshold settings
     in the RF. 

     Table~\ref{tab:trigger} shows the estimated fraction of events that also record an RF signal,
    ranging from about 40\% at $\SI{10}{PeV}$ to above 90\% at $\SI{100}{PeV}$.
    The total number of events with RF content would be in the range of 5 to 10 events.

        
    \begin{table}
      \centering
      \begin{tabular}{|c||c|c|c|}
        \hline
        Energy &  Fraction of events with RF content \\
        \hline
        \hline
        $\SI{6}{PeV}$ &  0.11  \\
        $\SI{10}{PeV}$ &  0.41  \\
        $\SI{30}{PeV}$ &  0.84   \\
        $\SI{100}{PeV}$  &  0.91  \\
        \hline
      \end{tabular}
      \caption{The  fraction of contained events with RF content are shown.  
      The requirement for the RF channel is at least 2 sigma on 2 antennas. }
      \label{tab:trigger}
    \end{table}

The option to deploy antenna arrays with self trigger
capability was not investigated.  
The implementation approach would be very different.  Phased arrays of antennas would be installed primarily on the outer strings of the array, 
and primarily at shallower depth above $\SI{1500}{m}$ where the coldest ice is located. 
An initial estimate indicates that the effective volume at very high energies (> $\SI{1}{EeV}$) would be substantially larger.  
Such an approach would be intriguing but the technical realization 
would be a sizeable project, despite the "free" boreholes.

\section{Summary}

At the highest energies due to Earth absorption most of the arriving neutrinos are from the southern sky, making it difficult to reject cosmic-ray background except for starting events. The cascade channel is the main contributor to probe Universe beyond multi-PeVs. A study was performed of  a Gen2 style detector configuration
that is augmented with vertically polarized antennas,
one antenna connected to every optical sensor.  
A complete hybrid RF/optical simulation was performed of such volume detector. 
At energies above a few PeV the events are starting to contain 
some RF signals, at $\SI{10}{PeV}$ the efficiency is about 50\%. 
Even a very coarse attempt to use the RF information for reconstruction
shows that the directional reconstruction could improve substantially to the
range of 1 degree and possibly well below. 
For the simulations only contained events were used. The addition of events with 
vertex outside the array but still with a bright optical signal (> 1000 photoelectrons)  may increase the effective volume slightly. 
In addition to the multi-messenger opportunities that would be made available by embedding radio antennas in the IceCube detector, the additional radio data may also benefit signal/background separation of non-contained events pointing towards the detector.
  However the overall rates with the chosen configuration are modest.
For the choice of antenna, events below $\SI{3}{PeV}$ would rarely 
have an RF signal.
The option to deploy antenna arrays with self trigger
capability was not investigated. An initial estimate indicates that the effective volume at very high energies (> $\SI{1}{EeV}$) would be an order of magnitude larger. However it would be 
still short of the science goals of Gen2 and be significantly smaller than
the planned Gen2 radio array.





\bibliographystyle{ICRC}
\bibliography{references}

\providecommand{\href}[2]{#2}\begingroup\raggedright\begin{thebibliography}{10}

\bibitem{Gen2}
{\bfseries IceCube} Collaboration
  \href{http://dx.doi.org/10.1088/1361-6471/abbd48}{{\em Journal of Physics G:
  Nuclear and Particle Physics} {\bfseries 48} no.~6, (Apr, 2021) 060501}.

\bibitem{ICFlux1}
{\bfseries IceCube} Collaboration
  \href{http://dx.doi.org/10.1126/science.1242856}{{\em Science} {\bfseries
  342} no.~6161, (2013) }.

\bibitem{ICFlux2}
{\bfseries IceCube} Collaboration
  \href{http://dx.doi.org/10.3847/0004-637x/833/1/3}{{\em The Astrophysical
  Journal} {\bfseries 833} no.~1, (Dec, 2016) 3}.

\bibitem{ICFlux3}
{\bfseries IceCube} Collaboration
  \href{http://dx.doi.org/10.1103/PhysRevLett.125.121104}{{\em Phys. Rev.
  Lett.} {\bfseries 125} (Sep, 2020) 121104}.

\bibitem{ICFlux4}
{\bfseries IceCube} Collaboration
  \href{http://dx.doi.org/10.1038/s41586-021-03256-1}{{\em Nature} {\bfseries
  591} no.~7849, (Mar, 2021) 220--224}.

\bibitem{RICE}
I.~Kravchenko, C.~Cooley, S.~Hussain, and et~al.
  \href{http://dx.doi.org/10.1103/physrevd.73.082002}{{\em Physical Review D}
  {\bfseries 73} no.~8, (Apr, 2006) }.

\bibitem{ANITA}
P.~Gorham, P.~Allison, O.~Banerjee, and et~al.
  \href{http://dx.doi.org/10.1103/physrevd.98.022001}{{\em Physical Review D}
  {\bfseries 98} no.~2, (Jul, 2018) }.

\bibitem{ARA}
{\bfseries ARA} Collaboration
  \href{http://dx.doi.org/10.1016/j.astropartphys.2011.11.010}{{\em
  Astroparticle Physics} {\bfseries 35} no.~7, (Feb., 2012) 457--477}.

\bibitem{ARA2}
{\bfseries ARA} Collaboration
  \href{http://dx.doi.org/10.1103/physrevd.102.043021}{{\em Physical Review D}
  {\bfseries 102} no.~4, (Aug, 2020) }.

\bibitem{ARIANNA1}
A.~Anker, S.~Barwick, H.~Bernhoff, and et~al.
  \href{http://dx.doi.org/10.1088/1475-7516/2020/03/053}{{\em Journal of
  Cosmology and Astroparticle Physics} {\bfseries 2020} no.~03, (Mar, 2020)
  053–053}.

\bibitem{ARIANNA2}
A.~Anker, S.~Barwick, H.~Bernhoff, and et~al.
  \href{http://dx.doi.org/10.1016/j.asr.2019.06.016}{{\em Advances in Space
  Research} {\bfseries 64} no.~12, (Dec, 2019) 2595–2609}.

\bibitem{RNOG}
J.~Aguilar, P.~Allison, J.~Beatty, and et~al.
  \href{http://dx.doi.org/10.1088/1748-0221/16/03/p03025}{{\em Journal of
  Instrumentation} {\bfseries 16} no.~03, (Mar, 2021) P03025}.

\bibitem{AURA}
H.~Landsman, L.~Ruckman, and G.~Varner
  \href{http://dx.doi.org/https://doi.org/10.1016/j.nima.2009.03.030}{{\em
  Nuclear Instruments and Methods in Physics Research Section A: Accelerators,
  Spectrometers, Detectors and Associated Equipment} {\bfseries 604} no.~1,
  Supplement, (2009) S70--S75}. ARENA 2008.

\bibitem{CTW-interaction}
A.~Connolly, R.~S. Thorne, and D.~Waters
  \href{http://dx.doi.org/10.1103/physrevd.83.113009}{{\em Physical Review D}
  {\bfseries 83} no.~11, (Jun, 2011) }.

\bibitem{Gandhi}
A.~Bhattacharya, R.~Gandhi, W.~Rodejohann, and A.~Watanabe
  \href{http://dx.doi.org/10.1088/1475-7516/2011/10/017}{{\em Journal of
  Cosmology and Astroparticle Physics} {\bfseries 2011} no.~10, (Oct, 2011)
  017--017}.

\bibitem{ARZ-signal}
J.~Alvarez-Muñiz, A.~Romero-Wolf, and E.~Zas
  \href{http://dx.doi.org/10.1103/physrevd.84.103003}{{\em Physical Review D}
  {\bfseries 84} no.~10, (Nov, 2011) }.

\bibitem{ICRC_LOM}
{\bfseries IceCube Gen2} Collaboration {\em PoS} {\bfseries ICRC2021} (these
  proceedings) 1062.

\end{thebibliography}\endgroup


\providecommand{\href}[2]{#2}\begingroup\raggedright\endgroup



\clearpage
\section*{Full Author List: IceCube-Gen2 Collaboration}




\scriptsize
\noindent
R. Abbasi$^{17}$,
M. Ackermann$^{71}$,
J. Adams$^{22}$,
J. A. Aguilar$^{12}$,
M. Ahlers$^{26}$,
M. Ahrens$^{60}$,
C. Alispach$^{32}$,
P. Allison$^{24,\: 25}$,
A. A. Alves Jr.$^{35}$,
N. M. Amin$^{50}$
% ...
\\
\\
% ...
$^{12}$ Universit{\'e} Libre de Bruxelles, Science Faculty CP230, B-1050 Brussels, Belgium \\
% ...
$^{17}$ Department of Physics, Loyola University Chicago, Chicago, IL 60660, USA \\
% ...
$^{22}$ Dept. of Physics and Astronomy, University of Canterbury, Private Bag 4800, Christchurch, New Zealand \\
% ...
$^{24}$ Dept. of Astronomy, Ohio State University, Columbus, OH 43210, USA \\
$^{25}$ Dept. of Physics and Center for Cosmology and Astro-Particle Physics, Ohio State University, Columbus, OH 43210, USA \\
$^{26}$ Niels Bohr Institute, University of Copenhagen, DK-2100 Copenhagen, Denmark \\
% ...
$^{32}$ D{\'e}partement de physique nucl{\'e}aire et corpusculaire, Universit{\'e} de Gen{\`e}ve, CH-1211 Gen{\`e}ve, Switzerland \\
% ...
$^{35}$ Karlsruhe Institute of Technology, Institute for Astroparticle Physics, D-76021 Karlsruhe, Germany  \\
% ...
$^{50}$ Bartol Research Institute and Dept. of Physics and Astronomy, University of Delaware, Newark, DE 19716, USA \\
% ...
$^{60}$ Oskar Klein Centre and Dept. of Physics, Stockholm University, SE-10691 Stockholm, Sweden \\
% ...
$^{71}$ DESY, D-15738 Zeuthen, Germany \\
% ...

\subsection*{Acknowledgements}

\noindent
USA {\textendash} U.S. National Science Foundation-Office of Polar Programs,
U.S. National Science Foundation-Physics Division,
U.S. National Science Foundation-EPSCoR,
Wisconsin Alumni Research Foundation,
Center for High Throughput Computing (CHTC) at the University of Wisconsin{\textendash}Madison,
Open Science Grid (OSG),
Extreme Science and Engineering Discovery Environment (XSEDE),
Frontera computing project at the Texas Advanced Computing Center,
U.S. Department of Energy-National Energy Research Scientific Computing Center,
Particle astrophysics research computing center at the University of Maryland,
Institute for Cyber-Enabled Research at Michigan State University,
and Astroparticle physics computational facility at Marquette University;
Belgium {\textendash} Funds for Scientific Research (FRS-FNRS and FWO),
FWO Odysseus and Big Science programmes,
and Belgian Federal Science Policy Office (Belspo);
Germany {\textendash} Bundesministerium f{\"u}r Bildung und Forschung (BMBF),
Deutsche Forschungsgemeinschaft (DFG),
Helmholtz Alliance for Astroparticle Physics (HAP),
Initiative and Networking Fund of the Helmholtz Association,
Deutsches Elektronen Synchrotron (DESY),
and High Performance Computing cluster of the RWTH Aachen;
Sweden {\textendash} Swedish Research Council,
Swedish Polar Research Secretariat,
Swedish National Infrastructure for Computing (SNIC),
and Knut and Alice Wallenberg Foundation;
Australia {\textendash} Australian Research Council;
Canada {\textendash} Natural Sciences and Engineering Research Council of Canada,
Calcul Qu{\'e}bec, Compute Ontario, Canada Foundation for Innovation, WestGrid, and Compute Canada;
Denmark {\textendash} Villum Fonden and Carlsberg Foundation;
New Zealand {\textendash} Marsden Fund;
Japan {\textendash} Japan Society for Promotion of Science (JSPS)
and Institute for Global Prominent Research (IGPR) of Chiba University;
Korea {\textendash} National Research Foundation of Korea (NRF);
Switzerland {\textendash} Swiss National Science Foundation (SNSF);
United Kingdom {\textendash} Department of Physics, University of Oxford.

\end{document}